# Multi-BERT for Embeddings for Recommendation System

Krutika Sarode
Computer Science
University of Massachusetts Amherst
Amherst MA US
ksarode@umass.edu

Shashidhar Reddy Javaji
Computer Science
University of Massachusetts Amherst
Amherst MA US
sjavaji@umass.edu

## ABSTRACT

In this paper, we propose a novel approach for generating document embeddings using a combination of Sentence-BERT (SBERT) and RoBERTa, two state-of-the-art natural language processing models. Our approach treats sentences as tokens and generates embeddings for them, allowing the model to capture both intra-sentence and inter-sentence relations within a document. We evaluate our model on a book recommendation task and demonstrate its effectiveness in generating more semantically rich and accurate document embeddings. To assess the performance of our approach, we conducted experiments on a book recommendation task using the Goodreads dataset. We compared the document embeddings generated using our MULTI-BERT model to those generated using SBERT alone. We used precision as our evaluation metric to compare the quality of the generated embeddings. Our results showed that our model consistently outperformed SBERT in terms of the quality of the generated embeddings. Furthermore, we found that our model was able to capture more nuanced semantic relations within documents, leading to more accurate recommendations. Overall, our results demonstrate the effectiveness of our approach and suggest that it is a promising direction for improving the performance of recommendation systems

## INTRODUCTION

A subtype of information filtering system called a recommendation system makes suggestions for items that are most relevant to a certain user or query. Usually, the recommendations are made about various decision-making procedures, like choosing a product to buy, movies to watch, or online books to read. When a person must select an item from a service's possible overwhelming selection of things, recommender systems are especially helpful. There are two types of recommendation systems**[1]**, collaborative which mostly is user depend, it relies on the history and past actions of users and recommends based on those, the other is content-based which is dependent on content like title, description, genre, author, etc and also user's preference, this can be used when you have the information of the items and not the user, this paper works on content-based.

The recommender system that uses content has a number of benefits. First, as it is based on item representation, the content-based recommendation is user-independent. Therefore, this type of system is not affected by the data sparsity issue. Secondly, content-based recommender systems can address the new item cold-start issue by recommending new products to consumers. Last but not least, content-based recommender systems are able to explain the recommendation outcome in detail. In comparison to other methods, this kind of system's transparency has many advantages in real-world applications.

Many methods are used to perform content-based filtering, it includes TF-IDF**[2]**, vector space models, and classification, the basic idea of all these later models was to project the words into a high-dimensional space and then try to find the distance between them using techniques like cosine similarity which is finding how similar two orthogonal vectors are using the angle between them, Pearson correlation coefficient which is very similar to the previous one, euclidean distance, the most basic, apart from these clustering methods can also be applied to get the closest vector given a vector. Cosine similarity and clustering methods are used in this paper.

One big question is how are words projected into dimensional space. There are models like Bag-of-Words(BoW), This approach assigns a unique token, usually a number, to each word that appears in the text, Word2Vec, In many ways, Word2Vec builds on BoW but instead of assigning discrete tokens to words it learns continuous multi-dimensional vector representation for each word in the training corpus, GloVe **[3]**, etc. These models take words as input and convert them into vectors**[4,5]**, these vectors can be projected into higher dimensions and used for similarity calculation. The best among all these is Word2Vec, but its primary flaw is that it only offers one representation for a word that is the same no matter what context it is used in.

The solution to this problem was contextual word embeddings. The fundamental idea underlying contextual word embeddings is to prove more than one representation for each word depending on the context in which it appears. The emergence of contextual word embeddings was driven by the success of deep learning models in natural language processing tasks**[6]**. These models, such as recurrent neural networks (RNNs) **[7,8]** and convolutional neural networks (CNNs), require input text to be represented as a sequence of vectors, rather than a string of words. Traditional word embeddings, such as word2vec and GloVe, provided a way to convert words into vectors, but they did not take into account the context in which the words appeared.To address this limitation, researchers began developing contextual word embedding models, such as BERT**[9,10]** and ELMo**[11]**, which are able to capture the context in which a word appears and produce a vector representation that reflects this context. These models use a transformer architecture, which allows them to capture complex relationships between words in the input text, and they are trained on large corpora of text, which allows them to learn to produce accurate and context-aware word embeddings.

Recently there has also been more improvement with SBERT**[12,13,14]** being most popular for sentence-level tasks, SBERT was introduced as a way to improve the performance of BERT on sentence-level tasks, Unlike BERT (Bidirectional Encoder Representations from Transformers), which processes entire sequences of text, SBERT processes individual sentences and learns to represent each sentence in a fixed-length vector. This allows SBERT to better capture the meaning of sentences and improve the performance of NLP systems that operate at the sentence level.

But, there are limitations, SBERT processes the entire document as a single sentence. This is because SBERT is a sentence-level model, which means that it is designed to understand the meaning of individual sentences rather than individual words. By processing the entire document as a single sentence, SBERT is able to capture the overall meaning of the document, which is more generalized. To better this, this paper introduces MULTI-BERT which gives more power to SBERT by adding sentence embeddings created by RoBERTa**[15]**. Which would allow for more information capture.

## PROPOSED METHODOLOGY

It is a very simple model with few additions to the existing SBERT model, instead of just considering the whole document for embeddings, the sentences are taken and passed through RoBERTa model, The MULTI-BERT technique involves splitting a document into its individual sentences, extracting the sentence embeddings using SBERT, and treating each sentence as a token. These sequences of tokens are then fed into an pretrained RoBERTa model**[16]**, and the last input layer of RoBERTa is extracted and used as the document vector. This document vector captures the inter-sentence relations within the document, and it is concatenated with the sentence embeddings to create the final document embedding.

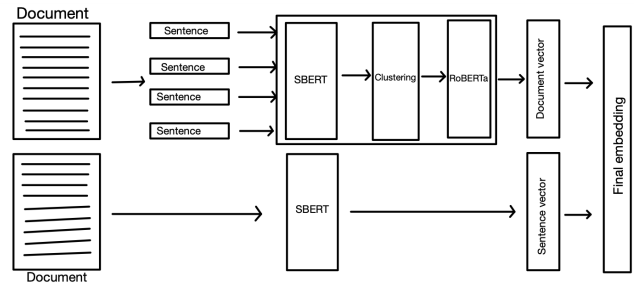

Figure 1: **Architecture of the Proposed Model**

We first divide the input document into individual sentences and generate sentence embeddings for each sentence using the SBERT model. This allows us to capture the contextual meaning of each sentence without exceeding the maximum length of the SBERT model. We then apply k-means clustering to these sentence embeddings, assigning each sentence to a cluster. We use 200 clusters in our experiments. This creates a "cluster codebook" that maps each sentence embedding to a cluster. We use this cluster codebook to create a document vector, where each sentence in the document is represented by its corresponding cluster id. We fine-tune the RoBERTa model by providing it with this document vector as input and the same vector as the output label. This allows the model to learn to generate document vectors that capture the semantic relations between sentences within a document. We use a batch size of 16, attention heads of 12, 6 hidden layers, and a position vocab size of 514. We train the model using the Adam optimizer with a learning rate of 1e-4. After training, we save the fine-tuned model for use in generating document embeddings. Our approach allows us to create more semantically rich and accurate document embeddings by combining the strengths of SBERT and RoBERTa[17]. By using sentence embeddings and clustering, we can capture the contextual meaning of each sentence and the relationships between sentences within a document. This allows us to better identify similar books and make more accurate recommendations.

Once we have trained and saved our fine-tuned RoBERTa model, we can use it to generate document embeddings for any input document. To do this, we follow the same process as before: we divide the input document into individual sentences, generate sentence embeddings using SBERT, assign each sentence to a cluster using the cluster codebook, and create a document vector based on the cluster id of each sentence. We then provide this document vector to the fine-tuned RoBERTa model, which generates a corresponding document embedding. This document embedding captures the contextual meaning of each sentence within the document, as well as the relationships between sentences. We can use these document embeddings in a variety of ways, such as to make recommendations or to perform similarity calculations between documents. For example, to make a recommendation, we could generate document embeddings for a set of books and use a similarity measure such as cosine similarity to identify the most similar books to a given input book. This would allow us to recommend books that are similar in terms of their content and context, rather than just their metadata.

While cosine similarity is a commonly used measure of similarity between vectors, it has some limitations when applied to document embeddings. Cosine similarity measures the angle between two vectors, which is useful for identifying similar vectors but may not always be the best approach for document embeddings. In contrast, k-means clustering[18] is a technique that groups similar vectors together and can be more effective for document embeddings because it takes into account the relationships between vectors within a cluster. Additionally, k-means clustering can handle larger and more complex datasets than cosine similarity, making it a better choice for many applications involving document embeddings. Furthermore, k-means clustering allows for the identification of clusters of similar documents, which can be useful for making more accurate recommendations.

$$\arg\min_{\mathbf{S}} \sum_{i=1}^{k} \sum_{\mathbf{x} \in S_i} \|\mathbf{x} - \boldsymbol{\mu}_i\|^2 = \arg\min_{\mathbf{S}} \sum_{i=1}^{k} |S_i| \operatorname{Var} S_i$$

For example, if a user is interested in a particular book, k-means clustering(above Equation) can identify other books that are similar to it and recommend those to the user. This is not possible with cosine similarity, which only measures the similarity between two individual vectors. In short, k-means clustering is a more powerful and flexible tool for working with document embeddings compared to cosine similarity and is better suited for many applications in the field of recommendation systems.

### Data

The Goodreads book reviews dataset provided by the University of California, San Diego includes detailed information on millions of books, ratings, and reviews from the Goodreads website. The dataset is organized into several tables, each containing specific information about the books and reviews. In this study, we focus on the children's genre, which includes 124,082 books and 734,640 detailed reviews.

The book's table includes information such as the title, author, and publication date, while the rating table includes the rating given to the book by each user along with the review text and other metadata. The user's table includes information about the users who submitted ratings and reviews, such as their name, location and reading history. The data is structured and allows for in-depth analysis.

### Baselines

The baseline models considered for this experiment are SBERT, and TF-IDF. SBERT is a transformer-based language model that has been trained to encode contextual information from sentences. It is a variant of BERT, which is a widely-used model for natural language processing tasks.

SBERT uses a two-sentence input representation, where each sentence is encoded using the BERT model. The encoded representations of the two sentences are then concatenated and passed through a classification layer to predict the relationship between the two sentences. This allows SBERT to capture contextual information from the sentences and use it for downstream tasks, such as semantic similarity, natural language inference, and sentiment analysis.

One of the main advantages of SBERT over other language models is its ability to capture sentence-level contextual information, which is important for many natural language processing tasks. Additionally, SBERT has been shown to perform well on a wide range of tasks and achieve state-of-the-art results on several benchmarks.

TF-IDF is calculated by multiplying two statistics: the term frequency and the inverse document frequency. The term frequency is a measure of how often a term appears in a document, while the inverse document frequency is a measure of how rare the term is across the entire corpus of documents.

In a book recommendation system, TF-IDF can be used to determine the relative importance of words in a given book, and recommend books that have similar patterns of word usage. For example, if a user has read a book that contains the words "cat" and "dog" frequently, the system might recommend other books that also use these words frequently.

TF-IDF can also be used to compare the content of two or more books and determine how similar or dissimilar they are in terms of their vocabulary and language usage. This can be useful for recommending books that are similar in content to a user's previously-read books.

## Experiment and Results

We experimented over Goodreads datasets in which we picked the section of children books and took around 25000 records from children books data and children books review data. The children books collection consists of information for the books such as book title, authors , description, popular shelves the user has placed that book in, average rating of the book, number of pages. From the children books datasets we extracted columns namely, language_code,popular_shelves,is_ebook,average_rating,description,authors, book id, rating count, title. From the children book review dataset we took columns like review text,ratings and number of votes.

First, the raw training data is preprocessed by selecting only the relevant columns and removing any noisy data. This ensures that the data is clean and can be easily used by the model. Two datasets are then combined using a common column, resulting in a merged dataset that contains a subset of the information from each of the original datasets. Any columns with a high number of NULL values are replaced with a default value to ensure that the data is complete and can be used by the model. Overall, this preprocessing step ensures that the data is clean and ready for use by the model.

We then pass this data through each model and get the results. For evaluation purposes, we label the dataset using genres and compare the genres of a given book with the genres of recommended books. Since a book can belong to multiple genres, we create a one-hot encoding of these genres and calculate the relevance score of each recommended book by comparing the one-hot encodings. If the recommended book and the input book belong to at least a predetermined number of genres, then the recommended book is considered relevant. We then compute the precision at 5, 10, and 25 for the relevant recommended books in the order they are received.

We are doing a baseline model comparison by running the same evaluation metrics on a simple sentence-embedded model, MULTI-BERT model and TF-IDF vectorizer. The way we are evaluating the model is by using precision, and to be able to use precision we need to know which retrieved documents are relevant to the query and which are not, for that purpose we have considered genre to be the label, but in out dataset we have multiple genres for each data case.For taking the relevance of the recommended book, we have compared the genres of input book with recommended books and if the genres are matching above a certain threshold we consider that book to be relevant. We have set the threshold of this experiment to be greater than 40%.

| Models | P@5 | P@10 | P@25 |
|---|---|---|---|
| **MULTI-BERT** | 0.9413 | 0.7889 | 0.7621 |
| **S-BERT** | 0.7563 | 0.7764 | 0.7294 |
| **TF-IDF** | 0.8164 | 0.8128 | 0.7877 |

Table 1 : **Precision of different models**

The table above shows the precision values of different models at 5, 10, and 25 retrieved documents. The proposed model, MULTI-BERT, outperforms SBERT and TF-IDF at 5 retrieved documents, but falls behind TF-IDF at 10 and 25 retrieved documents. The performance of these models is dependent on the specific data and task at hand. In this case, the small amount of data used may have impacted the results. Additionally, TF-IDF's ability to identify important keywords within a document may have given it an advantage in the book recommendation task. Further experimentation with MULTI-BERT may reveal its potential to outperform existing models in other tasks.

## Conclusion and Future Work

MULTI-BERT is able to outperform SBERT because it captures both intra-sentence and inter-sentence relations within a document. SBERT, on the other hand, only captures the content of individual sentences and does not take into account the relationships between sentences within a document.By capturing both intra-sentence and inter-sentence relations, MULTI-BERT is able to create more powerful and accurate document embeddings than SBERT, which only captures the content of individual sentences. This is why MULTI-BERT is able to outperform SBERT.

When comparing the results of using TF-IDF and MULTI-BERT for retrieval, it is observed that MULTI-BERT performs well for small retrieval, but not as well for larger retrieval. This may be due to the limitations of MULTI-BERT, which can be further studied in future research to improve its performance for larger retrieval tasks.

Future work includes using this model on larger datasets and also more tasks other than recommendations. Also, we would like to try more hybrid models where multiple BERT models are combined to perform a single task. We would like to see how the performance of the model changes with the addition of other models.